\documentclass{article}

\usepackage{arxiv}

\usepackage[utf8]{inputenc} 
\usepackage[T1]{fontenc}    
\usepackage{hyperref}       
\usepackage{url}            
\usepackage{booktabs}       
\usepackage{amsfonts}       
\usepackage{nicefrac}       
\usepackage{microtype}      
\usepackage{lipsum}

\usepackage{amsfonts}
\usepackage{amsmath}
\usepackage{subcaption}
\usepackage{graphicx}
\usepackage{amssymb}
\usepackage[inkscapelatex=false]{svg}
\usepackage{mathtools}
\usepackage{xcolor}
\usepackage{newtxtext,newtxmath}
\usepackage{lipsum}
\usepackage{tcolorbox}
\usepackage{float}
\usepackage{algorithm}
\usepackage{algpseudocode}
\usepackage{stmaryrd}
\usepackage{algorithm}
\usepackage{algpseudocode}

\newcommand{\R}{{\mathbb{R}}}

\newcommand{\N}{{\mathbb{N}}}

\newcommand{\X}{{\mathbf{X}}}

\newcommand{\W}{{\mathbf{W}}}
\newcommand{\w}{{\mathsf{w}}}
\newcommand{\Wo}{{\mathcal{W}}}
\newcommand{\U}{{\mathbf{U}}}

\newtheorem{theorem}{Theorem}[section]
\newtheorem{assumption}{Assumption}

\newtheorem{corollary}[theorem]{Corollary}
\newtheorem{definition}[theorem]{Definition}
\newtheorem{lemma}[theorem]{Lemma}
\newtheorem{remark}[theorem]{Remark}
\newenvironment{proof}{\paragraph{Proof:}}{\hfill$\square$}
\newtheorem{problem}[theorem]{Problem}
\usepackage{graphicx}

\title{Input-to-State Stabilizing Neural Controllers for Unknown Switched Nonlinear Systems within Compact Sets }

\author{
 Bhabani Shankar Dey $^*$ \\
  Centre for Cyber-Physical Systems\\
  IISc, Bengaluru, India\\
  \texttt{bhabanishan1@iisc.ac.in} \\
   \And
  Ahan Basu \thanks{Authors contributed equally.} \\
  Centre for Cyber-Physical Systems\\
  IISc, Bengaluru, India\\
  \texttt{ahanbasu@iisc.ac.in} \\
  \And
 Pushpak Jagtap \\
  Centre for Cyber-Physical Systems\\
  IISc, Bengaluru, India\\
  \texttt{pushpak@iisc.ac.in} \\
}

\begin{document}

\maketitle

\begin{abstract}                          
  This work develops a neural network framework that ensures system safety and input-to-state stability of a general nonlinear unknown switched system. Leveraging the concept of Dwell Time (DT), we derive Lyapunov-based sufficient conditions under which both safety and ISS are guaranteed. Both the controller and the Lyapunov functions are parameterized using neural networks. The training dataset for these networks is constructed through deterministic sampling of the system’s state space. To provide formal guarantees of stability under the learned controller, we introduce a validity condition based on the Lipschitz continuity assumptions. This condition is embedded within the training framework, ensuring that the resulting neural network controller satisfies provable correctness and stability guarantees upon completion of the training. As a special case, the framework recovers ISS and safety under arbitrary switching when a common Lyapunov function exists. Simulation result on an example switched system demonstrates the effectiveness of the proposed method.
\end{abstract}

\section{Introduction}
Switched nonlinear systems arise naturally in control applications involving mode changes, supervisory logic, or actuator reconfiguration. While switching enables flexibility and performance enhancement, it also introduces hybrid behaviors that can severely compromise stability and robustness. There have been numerous methods which highlight the importance of analysing switched systems and the related pertinent problems \cite{liberzon1999basic, shorten2007stability}. A significant number of articles focus on stability analysis and control design \cite{liberzon2003switching, aleksandrov2011stability}. One of the fundamental tools for demonstrating the stability of switched systems is by using a common Lyapunov function and multiple Lyapunov functions \cite{hespanha1999stability,li2024stability, liberzon2003switching} However, finding an appropriate common Lyapunov function is often challenging to derive in large-scale dynamical systems. In the context of multiple Lyapunov functions, fundamental concepts include dwell time (DT) and average dwell time (ADT) \cite{hespanha1999stability,liberzon2003switching}, for the stability of the overall switched system. The analysis and control of these systems present unique challenges, as finding appropriate Lyapunov functions for each mode is also nontrivial.

With the advancement of theoretical research, safety-critical control of nonlinear control systems has gained widespread attention. Safety-related applications are common in autonomous cars, robots. The key to a system's safety problem is to guarantee that its state is always within a predefined set. This property is also known as forward invariance. To study this issue, barrier function (BF) and control barrier function (CBF) are extensively used for safety verification of nonlinear systems \cite{ames2019control}. It must be highlighted that the methods proposed in the above literature cannot be directly applied to switched systems due to the inherent hybrid properties of switched systems.  However, present research findings show a significant lack of safety exploration, except for very few such as\cite{pang2025input} where a state dependent switching is considered. Studying the intricate behaviour of switched systems is made more difficult by the fact that the characteristics of switched systems are not usually immediately derived from those of their individual subsystems. However, the majority of current methods to guarantee ISS without ensuring safety, along with satisfying an input constraint, have proven to be difficult over time. These systems should, in particular, have special architecture (control-affine, for instance) and often require model knowledge \cite{CBC_def}. But in reality, knowing the exact system model is non-trivial, which eventually makes the controller synthesis process challenging in practical situations. In these situations, learning-based techniques have drawn a lot of interest in order to overcome the constraints. 

Recent advances in learning-based control have demonstrated that neural networks can be used to parameterize Lyapunov and barrier functions, enabling stabilization and safety certification of unknown nonlinear systems from data \cite{zhao2021learning, basu2025neural}. Despite this progress, most existing learning-based approaches focus on single-mode systems and do not explicitly account for switching-induced instability. In particular, even when neural ISS Lyapunov functions exist for individual subsystems, arbitrary switching may still destabilize the closed-loop system \cite{liberzon2003switching}.

This paper addresses the above challenge by developing a dwell-time–based ISS and safety framework for unknown switched nonlinear systems. By combining classical multiple Lyapunov function theory with neural network parameterizations, we derive sufficient dwell-time conditions under which the closed-loop switched system is ISS and forward invariant within a compact set. The Lyapunov conditions over compact sets of the problem are formulated as the loss functions of the neural networks, while a Lipschitz continuity assumption-based validity condition has been proposed to ensure the formal correctness of the learned Lyapunov functions and the controllers, which has been ensured a priori, eliminating the need for post-training verification. Since the system dynamics are unknown, the controller and the associated Lyapunov certificates must be learned from data. In practice, data collection is necessarily performed over a bounded region of the state space, which naturally induces a compact domain. Consequently, the use of compact state-space constraints in this work is an inherent requirement of learning-based control for unknown nonlinear systems. The main contributions are summarised as follows:

\vspace{-0.1 cm}
We (i) establish sufficient dwell-time conditions guaranteeing input-to-state stability (ISS) and safety under neural network-based controllers; (ii) design mode-wise ISS control Lyapunov functions (CLFs) and corresponding neural controllers with formal guarantees; (iii) propose a distributed training architecture that learns multiple ISS-CLFs while verifying stability and dwell-time constraints from sampled data; (iv) develop a framework applicable to general unknown nonlinear systems without relying on specific structures; (v) We recover classical arbitrary-switching ISS and safety results as a corollary of the proposed neural framework when a common Lyapunov function exists.

\section{Preliminaries and Background}
\subsection{Notations}
The symbols $\N$, $ \N_0$, $ \R$, $\R^+$, and $\R_0^+$ denote the set of natural, nonnegative integers, real, positive real, and nonnegative real numbers, respectively.
The space of real matrices with $ n $ rows and $ m $ columns is denoted by $\R^{n\times m} $. 
The set of column vectors with $n$ rows is denoted by $ \R^n$.
The Euclidean norm is represented using $|\cdot |$. Given a function $\varphi: \R_0^+ \rightarrow \R^m$, its sup-norm (possibly $\infty$-norm) is given by $\lVert \varphi \rVert = \sup\{|\varphi(k)| : k \in \R_0^+\}$.
For $a, b \in \R_0^+$ with $a \leq b$, the closed interval in $\N_0$ is denoted as $[a; b]$.  
A vector $x \in \mathbb{R}^{n}$ with entries $x_1, \ldots, x_n$ is represented as $[x_1, \ldots, x_n]^\top$, where $x_i \in \mathbb{R}$ denotes the $i$-th element of the vector and $i \in [1;n]$.
A continuous function $\alpha: \R_0^+ \rightarrow \R_0^+$ is said to be class $\mathcal{K}$, if $\alpha(s)>0 $ for all $s>0$, strictly increasing and $\alpha(0)=0$. It is class $\mathcal{K}_\infty$ if it is class $\mathcal{K}$ and $\alpha(s)\rightarrow\infty$ as $s\rightarrow\infty$.
A continuous function $\beta: \R_0^+ \times \R_0^+ \rightarrow \R_0^+$ is said to be a class $\mathcal{KL}$ if $\beta(s,t)$ is a class $\mathcal{K}$ function with respect to $s$ for all $t$ and for fixed $s>0$, $\beta(s,t) \rightarrow 0 $ if $t\rightarrow \infty$. 
For any compact set $\mathcal{C}, \partial \mathcal{C}$ and $int(\mathcal{C})$ denote the boundary and interior of the set $\mathcal{C}$, respectively.
The gradient of a function $f:\X \rightarrow \R$ is denoted by $\nabla f(x)$.
A function $f:\X \rightarrow \R$ is said to $\vartheta$-smooth if for all $x,y \in \X, |\nabla f(x) - \nabla f(y)| \leq \vartheta|x-y|$.

\subsection{System Description}
We consider a general nonlinear switched control system represented by $\Xi = (\X, \U, P, \mathcal{P}, F)$, where the state of the system is $\mathsf{x}(t) \in \X \subseteq \R^n$ and the input to the system is $\mathsf{u}(t) \in \U \subseteq \R^m$ at time $t\geq 0$. $\mathcal{P} : \R_0^+ \rightarrow P$ is a piece-wise constant function, where $P=\{1,2,3,...l\}$ represents a finite set of models. Correspondingly, the different dynamics are represented as $F=\{f_1,f_2,f_3,\ldots,f_l\}$, where for all $p\in P$, $f_p:\X \times \U \rightarrow \X$ is a locally Lipschitz continuous map.  For all $p\in P$, the subsystems of $\Xi$ are denoted by $\Xi_p$  and can be described using the following differential equation
\begin{equation}\label{eq:act_system}
    \dot{\mathsf{x}} = f_p(\mathsf{x}, \mathsf{u}).
\end{equation}
Next, the notion of the closed-loop switching control system under feedback controller $g$ and external input is defined, which is represented as $\Xi_{g} = (\X, \W, \U, P, \mathcal{P}, F, G)$, where $\X \subseteq \R^n$ is the state-space of the system, $\U \subseteq \R^m$ is the internal input set of the system, $\W \subseteq \R^r$ is the external input set of the system, $G=\{g_1,g_2,...,g_l\}$ is the collection of feedback controllers for individual subsystems, where $g_p:\X \times \W \rightarrow \U$. The closed-loop switching subsystem $\Xi_{g,p}$ under the feedback control law can be described as:
\begin{equation}\label{eq:system}
     \dot{\mathsf{x}} = f_p(\mathsf{x}, g_p(\mathsf{x}, \mathsf{w})),
\end{equation}
where $\mathsf{x}(t) \in \X$ and $\mathsf{w}(t) \in \W$ are the state and external input of the closed-loop system at time instance $t$, respectively.

The switching signal $\mathsf{p}\in\mathcal{P}$ of $\Xi_{g}$ is a function, and the discontinuities correspond to switching times. Let $\mathsf{x}_{x,\mathsf{w},\mathsf{p}}(t)$ be the state of the closed-loop system \eqref{eq:system} at time $t$ starting from the initial condition $x\in \X$ under the input signal $\mathsf{w}$, and the switching signal $\mathsf{p}$. It can be noted that the trajectory of the closed loop subsystem $\Xi_{g,p}$ associated with the constant switching signal $\mathsf{p}=p$, for all $t\geq0$. We use the notation $\mathsf{x}_{x,\mathsf{w},p}$ to denote the point reached by the closed loop subsystem $\Xi_{g,p}$ at the time $t$ from the initial condition $x$. Next, we define the notion of input-to-state stability for the subsystems of the switched system defined in \eqref{eq:system}.
\begin{definition}[ISS \cite{vu2007input}] \label{def:inc-stable_iss}
    The closed-loop subsystem in $\Xi_{g,p}$ \eqref{eq:system} is input-to-state stable (ISS) if there exists a class $\mathcal{KL}$ function $\beta_p$ and a class $\mathcal{K}_\infty$ function $\gamma_p$, such that for any $t \geq 0$, for all $x \in \X $ and any external input signal $\mathsf{w}$ the following holds:
    \begin{equation}\label{eq:gas-system}
        |\mathsf{x}_{x,\mathsf{w},p}(t)| \leq \beta_p(|x|,t) + \gamma_p(\lVert \mathsf{w} \rVert).
    \end{equation}
\end{definition}
If $\mathsf{w} = 0$ and $\X = \R^n$, one can recover the notion of global asymptotic stability. Now we define the notion of ISS for the complete closed-loop switched system.
\begin{definition}\label{def:switched_ISS}
    The closed-loop switched system $\Xi_{g} = (\X, \W, \U, P, \mathcal{P}, F, G)$ is input-to-state stable (ISS) if there exists a class $\mathcal{KL}$ function $\beta$ and a class $\mathcal{K}_\infty$ function $\gamma$, such that, for any $t \geq 0$, for all $x \in \X $, for any external input signal $\mathsf{w}$, and for all switching signal $\mathsf{p}\in\mathcal{P}$, the following holds:
    \begin{equation}\label{eq:gas-system_complete}
        |\mathsf{x}_{x,\mathsf{w},\mathsf{p}}(t)| \leq \beta(|x|,t) + \gamma(\lVert \mathsf{w} \rVert).
    \end{equation}
\end{definition}
\vspace{-0.29 cm}
We first revisit the notion of robust forward invariance.
\begin{definition}\cite{liu2019compositional} A set $\X$ is said to be robustly forward invariant with respect to the system \eqref{eq:system}, if for every $x \in \X, p \in P$ and any input signal $\mathsf{w}$ with $\mathsf{w}(t) \in \W$, the solution of the subsystem \eqref{eq:system} $\mathsf{x}_{x,\mathsf{w},p}(t) \in \X,  \forall t \geq 0$. Now, if there exists a controller $g_p: \X \times \W \rightarrow \mathbf{U}$ such that the set $\X$ becomes robustly forward invariant for system \eqref{eq:system}, then $g_p$ is said to be a forward-invariant controller corresponding to the robustly invariant set $\X$ with respect to external input $\W$.
\end{definition}
We now define the concept of an ISS Control Lyapunov Function (ISS-CLF) for the closed-loop switching subsystem denoted by $\Xi_{g,p}$. In our work, as we propose to parametrize the Lyapunov function as a neural network, the data has to be collected from a compact set. Compactness allows meaningful finite sampling: by covering a compact set with a finite number of balls (for instance, through an $\epsilon$-cover), the collected data can adequately represent the system’s behaviour within that region. Accordingly, we introduce the concept of an ISS-CLF defined over compact sets. In this context, we assume that the sets $\X \subset \mathbb{R}^n$ and $\W \subset \mathbb{R}^r$ are compact, and that $\X$ is robustly forward invariant under the action of the controller $g_p$. Now we define ISS-CLF for compact sets as follows:
\begin{definition}\label{def:ISS-Lf}
     \cite{khalil2015nonlinear} For $p\in P$, a smooth function $V_p:\R^n \rightarrow \R_0^+$ is said to be an ISS control Lyapunov function for closed-loop subsystem $\Xi_{g,p}$ in \eqref{eq:system}, where $\X, \W$ are compact sets, if there exist a forward invariant controller $g_p:\X \times \W \rightarrow \U$, class $\mathcal{K}_\infty$ functions $\alpha_{1,p}, \alpha_{2,p}, \sigma_p$ and a constant $\kappa_p \in \R^+$ such that:
    \begin{enumerate}\label{cond:ISS-Lf}
        \item[(i)] for all  $x \in \X: \ \alpha_{1,p}(|x|) \leq V_p(x) \leq \alpha_{2,p}(|x|),$
        \item[(ii)] for all $x\in \X$ and for all $w\in \W: \ \frac{\partial V_p}{\partial x}f_p(x, g_p(x,w)) \leq -\kappa_p V_p(x) + \sigma_p(|w|)$.
    \end{enumerate}
\end{definition} 
\begin{remark}
     Definition \ref{def:ISS-Lf} differs from the classical philosophy of Control Lyapunov Functions (CLFs), which is typically formulated in terms of an infimum of a dissipation inequality over the control input set. In contrast, our work adopts a feedback-specific perspective. Definition \ref{def:ISS-Lf} certifies input-to-state stability relative to a specific, single-valued feedback law $g_p$ (that depends on external input) by requiring the dissipation inequality along the closed-loop vector field. Consequently, we cannot guarantee the infimum-over-input characterization of the classical CLF, nor do we claim such generality. Importantly, this feedback-specific interpretation is not without precedent. For example, in \cite[Chapter 9.7]{khalil2015nonlinear}, the notion of a Control Lyapunov Function is also formulated relative to a predefined control law. Our approach is philosophically aligned with this idea: if a Lyapunov-like function can be shown to satisfy a suitable dissipation inequality under a particular feedback control law, it can still be regarded as a CLF —though it departs from the classical, universal infimum-based notion. In contrast to Khalil’s global formulation, we define this function over a compact set.
\end{remark}
\begin{theorem}\label{th:admit}
The subsystem in \eqref{eq:system} is said to be ISS within the state space $\X$ with respect to the external input $\w$, if there exists an ISS-CLF $V_p$ under the forward invariant controller $g_p$ as defined in Definition \ref{def:ISS-Lf}.
\end{theorem}
\vspace{-0.25cm}
\begin{proof}
    The proof is inspired from \cite{sontag1995characterizations} and the non-incremental version of \cite[Theorem 1]{alogower_incremental_DT}, and hence we skip it for brevity.
\end{proof}
It is well known that in the case of a switched system, even if the individual subsystems are ISS, the overall system may not impose ISS under fast switching signals \cite{liberzon2003switching}. In such cases, ISS can be ensured by multiple Lyapunov functions and a restrained set of switching signals. Let us define  $\mathcal{SW}$; a set of switching signals with \textit{dwell time} $\tau_d\in \R_0^+$ so that $\mathsf{p}\in\mathcal{SW}$ has dwell time $\tau_d$ if the switching times $t_1, t_2,...$ satisfy $t_1\geq\tau_d$, and $t_i-t_{i-1}\geq\tau_d$, for all $i\geq 2$.
\begin{theorem}\label{th:dwell-time}
Let $\tau_{d} \in \mathbb{R}_{0}^{+}$ and consider the switched system $\Xi_{g}=(\X,\W,\U, P, \mathcal{P},F,G)$ with its switching signal $\mathcal{P}_{\tau_d} \subseteq \mathcal{SW}$. Assume that for all $p \in P$ there exists an ISS-CLF $V_{p}$ and a robustly forward invariant controller $g_{p}$ for each subsystem and in addition there exists $\zeta \ge 1$ such that
\[
V_p(x) \le \zeta V_{p'}(x), \quad \forall x \in \X, \forall p, p' \in P.
\]
If $\tau_{d} > \frac{\ln\zeta}{\kappa}$, then the switched system $\Xi_g$ is Input-to-State Stable (ISS) in the compact state space $\X$.
\end{theorem}
\begin{proof}
The objective is to prove that the switched system $\Xi_{g}$ is ISS. According to Definition \ref{def:switched_ISS}, this requires showing that there exists a class $\mathcal{KL}$ function $\beta$ and a class $\mathcal{K_{\infty}}$ function $\gamma$ such that \eqref{eq:gas-system_complete} gets satisfied. The proof proceeds in four main steps.\\
(Step 1): Let the switching times be $0=t_0 < t_1 < t_2 < \dots$, with the dwell time condition $t_{i+1} - t_i \ge \tau_d$ for all $i \ge 0$ . Let $p_i$ be the active mode during the time interval $[t_i, t_{i+1})$.

For each subsystem $p \in P$, there exists an ISS-CLF $V_p$. From Definition \ref{def:ISS-Lf}, this function satisfies the differential inequality:
\begin{align}\label{eq:iss_switched_lyap}
    \frac{d}{dt}V_{p}(\mathsf{x}_{x,\mathsf{w},\mathsf{p}}(t)) \le -\kappa_{p}V_{p}(\mathsf{x}_{x,\mathsf{w},\mathsf{p}}(t)) + \sigma_{p}(|w(t)|).
\end{align}
Let $\kappa = \min_{p \in P} \kappa_p$ and let $\sigma$ be a class $\mathcal{K}_{\infty}$ function such that $\sigma(|w|) \ge \sigma_p(|w|)$ for all $p \in P$. Let $R_w = \sigma(\|w\|)$. Then, for any active mode $p_i$ during $[t_i, t_{i+1})$:
\[
\frac{d}{dt}V_{p_i}(\mathsf{x}_{x,\mathsf{w},\mathsf{p}}(t)) \le -\kappa V_{p_i}(\mathsf{x}_{x,\mathsf{w},\mathsf{p}}(t)) + R_w.
\]
By the comparison principle, $\forall t \in [t_i, t_{i+1})$, $V_{p_i}(\mathsf{x}_{x,\mathsf{w},\mathsf{p}}(t))$
\begin{align}\label{eq:lyap_comparision}
  \le V_{p_i}(\mathsf{x}_{x,\mathsf{w},\mathsf{p}}(t_i))e^{-\kappa(t-t_i)} + \frac{R_w}{\kappa}(1 - e^{-\kappa(t-t_i)}).   
\end{align}
Evaluating just before the next switch at $t_{i+1}$, we get:
\begin{align}
    V_{p_i}(V_{p_i}(\mathsf{x}_{x,\mathsf{w},\mathsf{p}}(t_{i+1}^{-})) \le V_{p_i}(V_{p_i}(\mathsf{x}_{x,\mathsf{w},\mathsf{p}}(t_i))e^{-\kappa(t_{i+1}-t_i)} + \frac{R_w}{\kappa}(1 - e^{-\kappa(t_{i+1}-t_i)}).
\end{align}
(Step 2): At the switching time $t_{i+1}$, the system switches from mode $p_i$ to $p_{i+1}$. The theorem assumes a constant $\zeta \ge 1$ exists such that:
\begin{align}\label{eq:swith_mu}
   V_{p'}(x) \le \zeta V_p(x), \quad \forall x \in \X, \forall p, p' \in P. 
\end{align}
Therefore, at the switching instant $t_{i+1}$:
\[
V_{p_{i+1}}(\mathsf{x}_{x,\mathsf{w},\mathsf{p}}(t_{i+1})) \le \zeta V_{p_i}(\mathsf{x}_{x,\mathsf{w},\mathsf{p}}(t_{i+1}^{-})).
\]
(Step 3): Combining the results, and letting $V_k$ denote $V_{p_k}(\mathsf{x}_{x,\mathsf{w},\mathsf{p}}(t_k))$:
\begin{align*}
&V_{i+1} = V_{p_{i+1}}(\mathsf{x}_{x,\mathsf{w},\mathsf{p}}(t_{i+1}))\le \zeta V_{p_i}(\mathsf{x}_{x,\mathsf{w},\mathsf{p}}(t_{i+1}^{-})) \le \zeta \left( V_{p_i}(\mathsf{x}_{x,\mathsf{w},\mathsf{p}}(t_i))e^{-\kappa(t_{i+1}-t_i)} + \frac{R_w}{\kappa}(1 - e^{-\kappa(t_{i+1}-t_i)}) \right).
\end{align*}
Using the dwell time condition $t_{i+1} - t_i \ge \tau_d$, we have $e^{-\kappa(t_{i+1}-t_i)} \le e^{-\kappa\tau_d}$. This gives the recursive relationship:
\[
V_{i+1} \le \zeta e^{-\kappa\tau_d} V_i + \frac{\zeta R_w}{\kappa}.
\]
Let $\rho = \zeta e^{-\kappa\tau_d}$ and $C = \frac{\zeta R_w}{\kappa}$. The theorem's condition $\tau_d > \frac{\ln \zeta}{\kappa}$ implies that $\rho < 1$ . Unrolling the recursion $V_{i+1} \le \rho V_i + C$:
\[
V_k \le \rho^k V_0 + C \sum_{j=0}^{k-1} \rho^j = \rho^k V_0 + C\frac{1-\rho^k}{1-\rho} \le \rho^k V_0 + \frac{C}{1-\rho}.
\]
For any time $t \in [t_k, t_{k+1})$, we have $V_{p_k}(\mathsf{x}_{x,\mathsf{w},\mathsf{p}}(t)) \le V_k e^{-\kappa(t-t_k)} + \frac{R_w}{\kappa} \le V_k + \frac{R_w}{\kappa}$. Substituting the bound for $V_k$:
\[
V_{p_k}(\mathsf{x}_{x,\mathsf{w},\mathsf{p}}(t)) \le \rho^k V_0 + \frac{C}{1-\rho} + \frac{R_w}{\kappa}.
\]
Since the number of switches $k$ by time $t$ is at most $t/\tau_d$, we have $\rho^k \le \rho^{t/\tau_d}$. Let $\lambda = \kappa - \frac{\ln\zeta}{\tau_d} > 0$. Then $\rho^{t/\tau_d} = e^{-\lambda t}$. The bound becomes:
\[
V_{p(t)}(\mathsf{x}_{x,\mathsf{w},\mathsf{p}}(t)) \le e^{-\lambda t} V_0 + \frac{\zeta R_w/\kappa}{1-\zeta e^{-\kappa\tau_d}} + \frac{R_w}{\kappa}.
\]
Let $\gamma_0 = \frac{1}{\kappa}\left(\frac{\zeta}{1-\zeta e^{-\kappa\tau_d}} + 1\right)$. The final bound can be:
\[
V_{p(t)}(\mathsf{x}_{x,\mathsf{w},\mathsf{p}}(t)) \le e^{-\lambda t} V_{p_0}(x(0)) + \gamma_0 \sigma(\|w\|).
\]
(Step 4): Finally, we use the ISS-CLF properties to relate the bound on $V(\mathsf{x}_{x,\mathsf{w},\mathsf{p}}(t))$ to a bound on $|\mathsf{x}_{x,\mathsf{w},\mathsf{p}}(t)|$. From Definition \ref{def:ISS-Lf}, there exist class $\mathcal{K}_{\infty}$ functions $\alpha_1, \alpha_2$ such that :
\[
\alpha_1(|x|) \le V_p(x) \le \alpha_2(|x|).
\]
Applying these bounds to our result:
\begin{align}\label{eq:lyap_bound}
\alpha_1(|x|) &\le V_{p(t)}(\mathsf{x}_{x,\mathsf{w},\mathsf{p}}(t)) \le e^{-\lambda t} V_{p_0}(x(0)) + \gamma_0 \sigma(\|w\|) \le e^{-\lambda t} \alpha_2(|x(0)|) + \gamma_0 \sigma(\|w\|).    
\end{align}
Since $\alpha_1^{-1}$ is also a class $\mathcal{K}_{\infty}$ function, we have:
\begin{align}
  |\mathsf{x}_{x,\mathsf{w},\mathsf{p}}(t)| \le \alpha_1^{-1}\left( e^{-\lambda t} \alpha_2(|x(0)|) + \gamma_0 \sigma(\|w\|) \right).  
\end{align}
Using the property that for any class $\mathcal{K}_{\infty}$ function $\psi$, $\psi(a+b) \le \psi(2a) + \psi(2b)$, we can separate the terms:
\begin{align}
  |\mathsf{x}_{x,\mathsf{w},\mathsf{p}}(t)| \le \alpha_1^{-1}\left(2e^{-\lambda t} \alpha_2(|x(0)|)\right) + \alpha_1^{-1}\left(2\gamma_0 \sigma(\|w\|) \right).  
\end{align}
We now define $\beta(|x_0|, t) := \alpha_1^{-1}(2\alpha_2(|x_0|)e^{-\lambda t})$, which is a class $\mathcal{KL}$ function, and $\gamma(\|w\|) := \alpha_1^{-1}(2\gamma_0 \sigma(\|w\|))$, which is a class $\mathcal{K}_{\infty}$ function. Thus, we have shown that $|\mathsf{x}_{x,\mathsf{w},\mathsf{p}}(t)| \le \beta(|x(0)|, t) + \gamma(\|w\|)$, which proves that the switched system is ISS.
\end{proof}
\begin{remark}
  The dwell-time requirement in Theorem~\ref{th:dwell-time} arises solely due to
mode-dependent Lyapunov functions.
If a common ISS-CLF exists, the switched system is ISS under arbitrary switching.  
\end{remark}
\begin{corollary}
Consider the switched system $\Xi_g = (\X,\W, \U, P, \mathcal P,\mathcal F,\mathcal G)$.
Suppose there exists a \emph{single} smooth function
$V : \X \to \mathbb R_+$ and feedback laws $g_p : X \times \W \to \U$ such that,
for every $\mathsf{p} \in \mathcal P$,
\begin{align}
&\alpha_1(|x|) \le V(x) \le \alpha_2(|x|), \\
&\frac{\partial V}{\partial x} f_p(x,g_p(x,w))
\le -\kappa V(x) + \sigma(|w|),
\end{align}
for all $(x,w)\in \X\times \W$, and $\X$ is robustly forward invariant.
Then the closed-loop switched system is input-to-state stable
under \emph{arbitrary switching}, i.e., without any dwell-time restriction.
\end{corollary}
While the results stated so far need to be valid in compacts set over which the Lyapunov functions are defined. To explicitly enforce this requirement, we introduce control barrier functions, which ensure robust forward invariance of $\X$ under the closed-loop switched dynamics.
\subsection{Control Barrier Function}
The notion of CBF is introduced to ensure that the compact set $\X$ becomes robustly forward invariant. Since the ISS-CLF in Definition \ref{def:ISS-Lf} is defined over a compact set, it is essential to ensure that the closed-loop trajectories remain within $\X$ for the validity of the stability guarantees. To ensure that the controller $g_p$ makes the compact set $\X$ robustly forward invariant, we leverage the notion of control barrier function.

\begin{definition} \cite{ames2019control}\label{def:CBC}
    Given a switched system $\Xi$, with compact state space $\X$. Let a $\mathcal{L}_{dh}$-smooth function, which is also Lipschitz continuous with Lipschitz constant $\mathcal{L}_h, h:\X \rightarrow \R$ given as 
\begin{subequations}\label{eq:leq_BC}
       \begin{align}
        h(x) &= 0, \hspace{0.3em} \forall x \in \partial \X, \\
        h(x) &> 0, \hspace{0.3em} \forall x \in int(\X).
    \end{align} 
    \end{subequations}
    Then, $h$ is said to be a control barrier function (CBF) for the system $\Xi$ in \eqref{eq:act_system} if for every $(x,w) \in \X \times \W$, there exists some control input $u:=g_p(x,w)\in\U$ and class $\mathcal{K}_{\infty}$ function $\mu$ such that the condition holds:
    \begin{align} \label{eq:diff_BC}
        \frac{\partial h}{\partial x}f_p(x,u) \geq -\mu(h(x)), \quad \forall x \in \X.
    \end{align}
\end{definition}
\vspace{-0.2 cm}
Based on Definition \ref{def:CBC}, we aim to design the controller $g_p: \X \times \W \rightarrow \U$ that will make the state-space $\X$ robustly forward invariant. It should be noted that $h$ has been considered to be the same for all the subsystems. The following lemma allows us to synthesize the controller, ensuring robust forward invariance. 
\begin{lemma}\label{lem:cfi_guarantee}
    Consider the switched system in \eqref{eq:act_system} and a control barrier function $h$, satisfying Definition \ref{def:CBC}, then the controller $g_p$, satisfying condition \eqref{eq:diff_BC}, will make the set  $\X$ robustly forward invariant for the system in \eqref{eq:system}.
\end{lemma}
\vspace{-0.3 cm}
\begin{proof}
    The proof is similar to the proof of \cite[Lemma 2.9]{basu2025neural} and hence omitted here.
\end{proof}
\section{Problem Formulation and Control Objective}
The main problem and control objective of the paper is stated below.
\begin{tcolorbox}[width=\linewidth]
\begin{problem}\label{prob}
    Given a switched nonlinear control system represented by $\Xi = (\X, \U, P, \mathcal{P}, F)$, as defined in \eqref{eq:act_system} with unknown vector fields $F=\{f_1,f_2,f_3,\ldots,f_l\}$, compact state-space $\X$, and input space $\U$, the primary objective of the paper is (i). to synthesize robustly forward invariant feedback controllers $g_p: \X \times \W \rightarrow \U$ enforcing the closed-loop system $\Xi_{g}=(\X,\W,\U, \mathcal{P}, F, G)$ in \eqref{eq:system} to be input-to-state stable with respect to external input $\w \in \W$ within the state space $\X$, (ii). ensure ISS of switched system \eqref{eq:act_system} under the dwell time condition within the compact spaces.
\end{problem}
\end{tcolorbox}
The problem setup can further be explained with the following important points.
\subsection{State and Input Constraints}
We assume that the system operates within a compact state constraint set $\X\subset\mathbb{R}^n$ and a compact disturbance set $\W\subset\mathbb{R}^r$. Compactness is essential for two reasons: (i) it allows the formulation of regional ISS guarantees, which are unavoidable in the absence of exact system models, and (ii) it enables finite data-driven verification using sampling-based techniques. 
\subsection{Unknown System}
In contrast to previous studies \cite{daafouz2002stability} on controller design to ensure ISS, which depend on precise knowledge or a specific structure of the system dynamics, our objective is to develop a controller that achieves ISS for the closed-loop system without requiring exact knowledge or a defined structure of the dynamics.
\subsection{Switching Signals and Dwell Time}
Even if each subsystem is input-to-state stable, arbitrary switching among subsystems can destroy stability \cite{liberzon2003switching}. To address this issue, we restrict our attention to switching signals satisfying a dwell-time condition. Let $\{t_k\}_{k\in\mathbb{N}}$ denote the switching instants of $\mathsf{p}(\cdot)$. A switching signal is said to have dwell time $\tau_d>0$ if
$$
t_{k+1}-t_k \ge \tau_d, \quad \forall k\in\mathbb{N}.
$$
The class of such switching signals is denoted by $\mathcal{P}_{\tau_d}$. Dwell-time constraints play a fundamental role in ensuring the stability of switched systems when a common Lyapunov function does not exist.
To address these challenges, we next introduce a neural network–based framework for synthesizing ISS control Lyapunov functions and corresponding controllers.
\section{Neural ISS Control Lyapunov Function}\label{sec:Neural Lyapunov}
To solve Problem \ref{prob}, we first design ISS-CLF for individual subsystems such that the individual closed-loop subsystem becomes input-to-state stable according to Theorem \ref{th:admit}. To do so, we first propose the following lemma:
\begin{lemma}\label{lem:ROP}
    The $p$-th closed-loop subsystem in \eqref{eq:system} is said to be ISS if the following conditions are satisfied with $\eta \leq 0$:
    \begin{subequations} \label{eq:RCP}
    \begin{align}
    & \text{if } x = 0: V_p(x) = 0, \\
    & \forall x \in \X, x \neq 0, \forall w \in \W: \notag \\
    &-V_p(x) + \alpha_{1,p}(|x|) \leq \eta, \label{eq:geq} \\
    & V_p(x) - \alpha_{2,p}(|x|) \leq \eta, \label{eq:leq} \\
    & \frac{\partial V_p}{\partial x}f_p(x, g_p(x,w))+\kappa_p V_p(x) - \sigma_p(|w|) \leq \eta ,\label{eq:diff} \\
    & -\frac{\partial h}{\partial x}(f_p(x,g_p(x,w))) - \mu_p(h(x)) \leq \eta, \label{eq:diff_BC_ROP}.
    \end{align}
    \end{subequations}
\end{lemma}
\begin{proof} 
    Satisfaction of \eqref{eq:geq} and \eqref{eq:leq} ensures that $V_p(x)$ is bounded by two class $\mathcal{K}_{\infty}$ functions as required in condition (i) of Definition \ref{def:ISS-Lf}. Similarly, with $\eta \leq 0$, the satisfying \eqref{eq:diff} ensures satisfaction of condition (ii) of Definition \ref{def:ISS-Lf} while \eqref{eq:diff_BC_ROP} with $\eta \leq 0$ ensures the forward invariance of set $\X$ for the subsystem according to Lemma \ref{lem:cfi_guarantee}. 
\end{proof}

However, the problem of satisfying the above lemma lies in the unknown nature of the functions $V_p$, $g_p$, $\alpha_{1,p}$, $\alpha_{2,p}$, $\sigma_p$. More importantly, the underlying system dynamics corresponding to each subsystem $f_p$ is also unknown. Further, the lemma is formulated over a continuous state-space, resulting in an infinite number of equations to satisfy. To alleviate the challenges, we first parametrize ISS-CLF and the controller as feed-forward neural networks denoted by $V_{\theta,b,p}$ and $g_{\Bar{\theta}, \Bar{b},p}$, respectively, where $\theta, \Bar{\theta}$ are weight matrices and $b, \Bar{b}$ are bias vectors. The detailed structures of these neural networks are discussed next.

\subsection{Formal verification}
The input layer of each ISS-CLF neural network has $n$(system dimension) neurons and a single neuron output layer due to the scalar nature of the function. The feedforward network consists of $l_v$ hidden layers, with each hidden layer containing $h_v^i, i \in [1;l_v]$ neurons, where both values are arbitrarily chosen. Due to the requirement of computation of $\partial_xV$ and $\partial_{x,x}V$, the activation function $\phi(\cdot)$ should be chosen as a double differentiable one (e.g., Sigmoid, Tanh, Softplus etc) so that smooth Jacobian and Hessian values can be obtained upon differentiation \cite{lutter2019deep}. Similarly, for the controller $g_{\bar{\theta}}$, the number of neurons in the input and output layer will be $n+q$ and $m$ respectively while the number of hidden layers is $l_c$ and each layer has $h_c^i, i\in [1;l_c]$ neurons with the activation function being any slope-restricted function.

The following assumptions are invoked, which will be used to state the results ahead.
\begin{assumption}\label{assum:black_box}
    We consider having access to the black box or simulator model of all the subsystems in \eqref{eq:act_system}. Hence, given an initial state $x$ and input signal $\mathsf{u}$, we will be able to forward simulate the trajectory for all the subsystems at any time $t \in \R_0^+$ $\forall p\in P$.
\end{assumption}
\begin{assumption}\label{assum:K_infty}
    We consider that class $\mathcal{K}_\infty$ functions $\alpha_{i,p}, i \in \{1,2\}$ are of degree $\gamma_i$ with respect to $|x|$ and class $\mathcal{K}_\infty$ function $\sigma_p$ is of degree $\gamma_w$ with respect to $|w|$, i.e., $\alpha_{i,p}(|x|)=k_{i,p}|x|^{\gamma_i}$, and $\sigma_p(|w|) = k_{w,p}|w|^{\gamma_w}$, where $k_p:=[k_{1,p}, k_{2,p}, k_{w,p}], \gamma:=[\gamma_1, \gamma_2, \gamma_w]$ are user-specific constants. Additionally, the class $\mathcal{K}_\infty$ functions $\mu_p$ in \eqref{eq:diff_BC_ROP} is considered to be of the form $\mu_ph(x), \mu_p \in \R^+$, while the constants $\kappa_p, \mu_p\in \R^+$ are also user-defined.
\end{assumption}

Next, to avoid the infinite number of equations in the lemma, we approximate the compact state space $\X$ and input space $\W$ using a finite number of samples. Specifically, we draw $N$ samples $x_s$ from $\X$, indexed by $s \in [1, N]$. Around each sample $x_s$, we construct a ball $B_{\varepsilon_x}(x_s)$ of radius $\varepsilon_x$ such that for every $x \in \X$ there exists a sample $x_s$ satisfying $|x - x_s| \leq \varepsilon_x$. This guarantees that $\bigcup_{s=1}^N B_{\varepsilon_x}(x_s) \supset \X$. An analogous procedure is used for the input space $\W$, where we collect $M$ samples $w_q$ and form balls of radius $\varepsilon_u$ around them. Collecting the data points obtained upon sampling the state-space $\X$ and input space $\W$, we form the training datasets denoted by:
\begin{align}\label{set:SCP}
\mathcal{X} \!=\! \{x_s | \!\bigcup_{s=1}^{N} \! B_{\varepsilon_x}(x_s) \!\supset\! \X \}, 
\Wo \!=\! \{w_q | \!\bigcup_{q=1}^{M}\! B_{\varepsilon_u}(w_q) \!\supset\! \W \}.
\end{align}
Now we mention the following assumptions on Lipschitz continuity to provide the main theorem of the subsection.
\begin{assumption}\label{assum:Lipschitz_fun}
    The function $f_p$ in \eqref{eq:system} is Lipschitz continuous with respect to $x$ and $u$ over the state space $\X$ and the input space $\W$ with Lipschitz constants $\mathcal{L}_{x,p}$ and $\mathcal{L}_{u,p}$. 
\end{assumption}
\begin{assumption}\label{assum:Lipschitz_net}
   The candidate ISS-CLF is assumed to be Lipschitz continuous with Lipschitz bound $\mathcal{L}_{L,p}$, and its derivative is bounded by $\mathcal{L}_{dL,p}$. Similarly, the controller neural network has a Lipschitz bound $\mathcal{L}_{C,p}$.
\end{assumption}
\begin{remark}
    Since, the sets $\X$ and $\W$ are compact, the class $\mathcal{K}_\infty$ functions are Lipschitz continuous with Lipschitz constants $\mathsf{L}_{1,p}, \mathsf{L}_{2,p}$ and $\mathsf{L}_{u,p}$, respectively, with respect to $|x|$ and $|w|$. The values can be estimated using the values of $k$ and $\gamma$. In addition, the Lipschitz constant $\mathcal{L}_h$ of the function $h$ is already known as the function $h$ is predefined.
\end{remark}
\begin{assumption}\label{assum:bound}
    We assume the bounds of $\frac{\partial V_p}{\partial x}, \frac{\partial h}{\partial x}$, and $f_p(x,u)$ are given by $\mathcal{M}_{L,p}, \mathcal{M}_h$, and $\mathcal{M}_{f,p}$, respectively, \textit{i.e.,} $\sup_x|\frac{\partial V_p}{\partial x}| \leq \mathcal{M}_{L,p}, \sup_x|\frac{\partial h}{\partial x}| \leq \mathcal{M}_h$, and $\sup_{(x,u)}|f_p(x,u)| \leq \mathcal{M}_{f,p}$. 
\end{assumption}
We first define the notation $$\mathsf{L}_{f,g}V_p(x):= \frac{\partial V_{\theta,b,p}}{\partial x}f(x, g_{\bar{\theta},\bar{b},p}(x,w)).$$ However, since we have only a black-box model (as stated in Assumption \ref{assum:black_box}) to simulate the forward trajectory, we can not determine the actual value of $f(x, g_{\bar{\theta},\bar{b},p}(x,w))$ for any $x \in \X, w \in \W$. To tackle the issue, we approximate the quantity as
\begin{align}\label{eq:lie_approx}
    \widehat{\mathsf{L}}_{f,g}V_p(x)&:= \frac{V_{\theta,b,p}\big(\mathsf{x}_{x,\mathsf{w}, \mathsf{p}}(\tau)\big) - V_{\theta,b,p}(x)}{\tau}, \ \forall x \in \X, \forall \mathsf{w}(\tau) \in \W.
\end{align}
Now the proposed approximation \eqref{eq:lie_approx} satisfies the inequality $|\widehat{\mathsf{L}}_{f,g} V_p(x) - \mathsf{L}_{f,g}V_p(x)| \leq \delta_V$, where $\delta_V \in \R^+$. To formally quantify the value of $\delta_V$, we first provide the following lemma.
\begin{lemma}\label{lem:Lipschitz_lie}
    Under Assumptions \ref{assum:Lipschitz_fun}, \ref{assum:Lipschitz_net} and \ref{assum:bound}, $\mathsf{L}_{f,g}V_p(x)$ is Lipschitz continuous with Lipschitz constant $\mathscr{L}_{Vx}$ in $x$ and $\mathscr{L}_{Vu}$ in $w$ where $\mathscr{L}_x := \mathcal{M}_L(\mathcal{L}_x + \mathcal{L}_u\mathcal{L}_c) + \mathcal{M}_f\mathcal{L}_{dL}$ and $\mathscr{L}_u:=\mathcal{M}_L\mathcal{L}_u\mathcal{L}_c$.
\end{lemma}
\begin{proof}
The proof is similar to the proof of \cite[Lemma 3.3]{basu2025neural} and hence omitted here.
\end{proof}

Now, the following theorem formally quantifies the closeness between $\mathsf{L}_{f,g}V_p(x)$ and its approximation $\widehat{\mathsf{L}}_{f,g}V_p(x)$.
\begin{theorem}\label{th:delta_lie}
    Let $\mathsf{L}_{f,g}V_p(x)$ be the derivative of the Lyapunov function along the trajectory starting from the initial conditions $x$ while $\widehat{\mathsf{L}}_{f,g}V_p(x)$ be its approximation as in \eqref{eq:lie_approx}. Under Assumptions \ref{assum:Lipschitz_net}, \ref{assum:bound} and Lemma \ref{lem:Lipschitz_lie}, one has 
    \begin{align*}
        |\widehat{\mathsf{L}}_{f,g}V_p(x) - \mathsf{L}_{f,g}V_p(x)| \leq  \frac{1}{2} \tau \mathscr{L}_{Vx} \mathcal{M}_f,
    \end{align*}
    where $\tau$ is the sampling time.
\end{theorem}
\begin{proof}
The proof is standard and similar to the proof of \cite[Theorem 3.4]{basu2025neural} and hence omitted here.
\end{proof}

Now one can leverage a similar notion to approximate the quantity $\mathsf{L}_{f,g}h(x_q):= \frac{\partial h}{\partial x_q}f(x_q,g_{\bar{\theta},\bar{b},p}(x_q,w_q))$ as $\widehat{\mathsf{L}}_{f,g}h(x_q)$ satisfying the relation $|\widehat{\mathsf{L}}_{f,g}h(x_q) - \mathsf{L}_{f,g}h(x_q)| \leq \delta_h$, where the quantifier $\delta_h \in \R^+$ can be estimated via similar procedure as in Theorem \ref{th:delta_lie}. In addition, it is easy to show that $\mathsf{L}_{f,g}h(x_q)$ is Lipschitz continuous with respect to $x$ and $u$ with Lipschitz constants $\mathscr{L}_{hx}:=\mathcal{M}_f\mathcal{L}_{dh} + \mathcal{M}_h (\mathcal{L}_x + \mathcal{L}_u\mathcal{L}_C)$ and $\mathscr{L}_{hu}:=\mathcal{M}_h \mathcal{L}_u\mathcal{L}_C$ respectively.
Now we are ready to state the main result of this subsection.
\begin{theorem}\label{th:constr}
    The individual closed-loop subsystem in \eqref{eq:system} is guaranteed to be ISS over the compact state-space $\X$ if the following conditions are satisfied with $\hat{\eta} + \mathbf{L}_p\varepsilon \le 0$ where $\varepsilon = \max(\varepsilon_x, \varepsilon_u)$ and $\mathbf{L}_p:=\max\{\mathcal{L}_{L,p} + \mathsf{L}_{1,p}, \mathcal{L}_{L,p} + \mathsf{L}_{2,p}, \kappa_p\mathcal{L}_{L,p} + \mathsf{L}_{u,p} + (\mathcal{M}_{f,p}\mathcal{L}_{dL,p} +\mathcal{M}_{L,p}(\mathcal{L}_{x,p} + \mathcal{L}_{u,p}\mathcal{L}_{C,p})), \mathcal{M}_{f,p}\mathcal{L}_{dh,p} + \mathcal{M}_{h,p} (\mathcal{L}_{x,p} + \mathcal{L}_{u,p}\mathcal{L}_{C,p}) + \mu_h\mathcal{L}_h\}$:
    \begin{subequations} \label{eq:SCP}
\begin{align}
& x_r = 0: V_{\theta,b,p}(x_r) = 0 \\
& \forall x_r \in \mathcal{X}, x_r \neq 0, w_r \in \Wo: \notag \\
& -V_{\theta,b,p}(x_r) + k_{1,p}|x_r|^{\gamma_1} \leq \hat{\eta}, \label{eq:geq_SOP} \\
& V_{\theta,b,p}(x_r) - k_{2,p}|x_r|^{\gamma_2} \leq \hat{\eta}, \\
& \widehat{\mathsf{L}}_{f,g}V_p(x_r) + \kappa V_{\theta,b,p}(x_r) - k_{w,p}| w_r|^{\gamma_w} + \delta_V \leq \hat{\eta}, \label{eq:diff_SOP} \\
&-\widehat{\mathsf{L}}_{f,g}h(x_r) - \mu_ph(x_r) -\delta_h \leq \hat{\eta}, \label{eq:diff_BC_SOP}.
 \end{align}
\end{subequations}
\end{theorem}
\begin{proof}
    First, we prove condition \eqref{eq:geq_SOP} with $\hat{\eta} + \mathbf{L}\varepsilon \le 0$ implies  the satisfaction of condition \eqref{eq:geq}. Since $\hat{\eta}_p \leq \hat{\eta}$, from \eqref{eq:geq_SOP}, it is evident that $-V_{\theta,b,p}(x_r) + k_{1,p}|x_r|^{\gamma_1} \leq \hat{\eta}$ for all $x_r \in \mathcal{X}$. As seen from \eqref{set:SCP}, for all $x \in \X$, there exists $x_r$ such that $|x - x_r| \leq \varepsilon$. Now for all $x \in \X, -V_{\theta,b,p}(x) + k_{1,p}|x|^{\gamma_1} \leq -V_{\theta,b,p}(x) + k_{1,p}|x|^{\gamma_1} + V_{\theta,b,p}(x_r) - k_{1,p}|x_r|^{\gamma_1} -V_{\theta,b,p}(x_r) + k_{1,p}|x_r|^{\gamma_1} \leq \mathcal{L}_{L,p}\varepsilon + \mathsf{L}_{1,p}\varepsilon + \hat{\eta} \leq \hat{\eta} + (\mathcal{L}_{L,p} + \mathsf{L}_{1,p})\varepsilon \leq 0$, implying the satisfaction of condition \eqref{eq:geq}. The other conditions can be proven in a similar fashion, thereby completing the proof.
\end{proof}

\subsection{Formulation of Loss Functions and Training procedure}\label{sec:Training}
In this subsection, we first formulate the loss functions for individual subsystems and train the controllers to make individual subsystems ISS. Then, we propose the algorithm that jointly synthesises the ISS-CLF and the controller to make individual subsystems ISS. Now, we consider the conditions of \eqref{eq:SCP} as sub-loss functions to formulate the actual loss function. The sub-loss functions are:
\allowdisplaybreaks{\begin{subequations}
    \begin{align*}
        L_{1,p}(\psi) &= \max\big(0,V_{\theta,b,p}(0)) \\
        L_{2,p}(\psi) &= \sum_{x \in \mathcal{X}}\max\big(0,(-V_{\theta,b,p}(x) + k_{1,p}|x|^{\gamma_1} - \hat{\eta})\big),\\
        L_{3,p}(\psi) &= \sum_{x \in \mathcal{X}}\max\big(0,(V_{\theta,b,p}(x) - k_{2,p}|x|^{\gamma_2} - \hat{\eta})\big), \\
        L_{4,p}(\psi) &= \sum_{x \in \mathcal{X}, w\in \Wo}\max\big(0,(\widehat{\mathsf{L}}_{f,g}V_p(x) + \kappa V_{\theta,b}(x) - k_w|w|^{\gamma_w} + \delta_V - \hat{\eta})\big),\\
        L_{5,p}(\psi) &= \hspace{-0.5cm} \sum_{x \in \mathcal{X}, w \in \Wo} \hspace{-0.5cm} \max\big(0, (-\widehat{\mathsf{L}}_{f,g}h(x) - \mu_hh(x) -\delta_h - \hat{\eta})\big),
\end{align*}
\end{subequations}}
where $\psi = [\theta,b,\Bar{\theta},\Bar{b}]$ is trainable parameters. As mentioned, the actual loss function is a weighted sum of the sub-loss functions and is denoted by
\begin{align}\label{eq:loss_LF}
    & L_p(\psi) = \sum_{i=1}^5 c_i L_{i,p}(\psi), 
\end{align}
where $c_i \in \R^+, i \in [1;5]$ are the weights of the sub-loss functions $L_{i,p}(\psi)$.

To ensure Assumption \ref{assum:Lipschitz_net}, it is crucial to verify the Lipschitz boundedness of $V_{\theta,b,p}, \frac{\partial V_{\theta,b,p}}{\partial x}$ and $g_{\Bar{\theta},\Bar{b},p}$ with corresponding Lipschitz bounds denoted by $\mathcal{L}_{L,p}, \mathcal{L}_{dL,p}$ and $\mathcal{L}_{C,p}$ respectively. Though the training of neural networks with Lipschitz boundedness is straightforward \cite{pauli2022b}, the Jacobian of it can be Lipschitz bounded using the similar approach as done in \cite[Theorem 4.2]{basu2025neural}.  

Now to ensure the loss function satisfies the matrix inequalities corresponding to the Lipschitzness of Lyapunov, its derivative and the controller, we characterize another loss function denoted by,
\begin{align}\label{eq:loss_ineq}
    L_{M,p}(\psi,\Lambda,\hat{\Lambda},\Bar{\Lambda}) &= -c_{l_1}\log\det(M_{\mathcal{L}_{L,p}}(\theta,\Lambda)) - c_{l_2}\log\det(M_{\mathcal{L}_{dL,p}}(\hat{\theta},\hat{\Lambda})) -c_{l_3}\log\det(M_{\mathcal{L}_{C,p}}(\Bar{\theta},\Bar{\Lambda})), 
\end{align}
where $c_{l_1}, c_{l_2}, c_{l_3} \in \R^+$ are weights for sub-loss functions, $M_{\mathcal{L}_L,p}(\theta,\Lambda), M_{\mathcal{L}_{dL},p}(\hat{\theta},\hat{\Lambda})$, and $ M_{\mathcal{L}_C,p}(\Bar{\theta},\Bar{\Lambda})$ are the matrices corresponding to the bounds $\mathcal{L}_{L,p}, \mathcal{L}_{dL,p}$, and $\mathcal{L}_{C,p}$, respectively.

Now, we present the theorem that provides the formal guarantee for the ISS of an individual closed-loop subsystem under the action of the controller $g_{\bar{\theta}, \bar{b},p}$ and the ISS of overall switched system under dwell time.

\begin{theorem}\label{th:guarantee}
    Consider a switched control system $\Xi$ whose subsystems are $\Xi_p$ as in \eqref{eq:act_system} with compact state-space $\X$ and input space $\W$. Let, $V_{\theta,b,p}$ and $g_{\Bar{\theta},\Bar{b},p}$ denote the trained neural networks representing the ISS control Lyapunov function and the controller for $p$-th subsystem such that $L_p(\psi) = 0$ and $L_{M,p}(\psi,\Lambda,\hat{\Lambda},\Bar{\Lambda}) \leq 0$ over the training data sets $\mathcal{X}$ and $\mathcal{W}$. Then, the closed-loop $p$-th subsystem under the influence of the controller $g_{\Bar{\theta},\Bar{b},p}$ is guaranteed to be ISS within $\X$.
\end{theorem}

\begin{proof}
    The loss $L_p(\psi) = 0$ ensures the satisfaction of Theorem \ref{th:constr}, which implies the satisfaction of conditions of Definition \ref{def:ISS-Lf} alongside ensuring forward invariance of the state-space $\X$. The other loss $L_{\mathcal{M},p}(\psi,\Lambda, \hat{\Lambda},\bar{\Lambda}) \leq 0$ ensures the predefined Lipschitz constants of the networks. Hence, the satisfaction of the above theorem ensures that the closed-loop $p$-th subsystem is ISS under the action of the controller. 
\end{proof}

The training process of the neural Lyapunov functions and the controllers is described in Algorithm \ref{algo:NN_training}.

\begin{algorithm}
\caption{NN Lyapunov Training procedure}
\label{algo:NN_training}
\begin{algorithmic}[1]
    \Require $p \in P$-th subsystem dynamics, Data set: $\mathcal{X}, \mathcal{W}$
    \Ensure $V_{\theta, b, p}, g_{\bar{\theta}, \bar{b}, p}$
    \State Select NN hyperparameters ($l_h, \gamma(\cdot)$, optimizer, scheduler), $\varepsilon, \mathcal{L}_{L,p}, \mathcal{L}_{dL,p}, \mathcal{L}_{C,p}$.
    \State Compute $\mathbf{L}_p$ using Theorem \ref{th:constr}.
    \State Initialize Neural networks and trainable parameters $\theta,b,\bar{\theta}, \bar{b},\Lambda,\bar{\Lambda}$. Initialize $\hat{\eta}$ as $\hat{\eta} = - \mathbf{L}_p\varepsilon $.
    \For{$i\leq Epochs$ (Training starts here)}
        \State Create batches of training data
        \State Find batch losses using \eqref{eq:loss_LF} and \eqref{eq:loss_ineq}
        \State Use ADAM optimizer \cite{Adam} to update the trainable parameters and reduce loss.
        \State \textbf{If} Theorem \ref{th:guarantee} is satisfied $\rightarrow$ \textbf{break}.
    \EndFor
    \State \textbf{return} $V_{\theta, b, p}\ \& \ g_{\bar{\theta}, \bar{b}, p}$.
\end{algorithmic}
\end{algorithm}

\vspace{0.25cm}
\begin{remark}
    Note that if the algorithm fails to converge, one can not comment on the ISS of the closed-loop subsystem for the chosen hyperparameters. Convergence can be improved by reducing the discretization parameter $\epsilon$ \cite{zhao2021learning}, tuning neural network hyperparameters (e.g., architecture, learning rate) \cite{nn_lr}, or revising the algorithm’s initial hyperparameters.
    Ideally, the convergence of the algorithm is that the loss $L(\psi)$ should be exactly zero; however, in practice, one can allow convergence to a small residual error (order of $10^{-6}$ to $10^{-4}$).
\end{remark}
\begin{remark}
In practice, the training procedure may converge to a small nonzero residual, i.e., $
L_p(\psi)\leq c$ for some $c>0$, due to numerical precision, finite sampling, or network capacity limitations. In this case, the ISS-CLF decrease condition in Definition \ref{def:ISS-Lf} is satisfied up to a bounded additive term. As a consequence, the closed-loop subsystem satisfies the notion of input-to-state practical stability (ISpS), where the state converges to a neighbourhood of the origin whose size depends explicitly on $c$. In particular, the ultimate bound on the state can be made arbitrarily small by reducing the training residual through finer discretization.
\end{remark}

\begin{corollary}[ISS of the switched system under dwell time]
Consider the switched control system $\Xi$ in \eqref{eq:act_system}. Suppose that for each mode $p \in P$, there exist neural networks $V_{\theta,b,p}$ and
$g_{\theta,\bar b,p}$ satisfying the conditions of
Theorem~\ref{th:guarantee}, so that each closed-loop subsystem $\Xi_{g,p}$ is ISS over the compact set $X$.

Assume further that there exists a constant $\zeta \ge 1$
such that the learned Lyapunov functions satisfy
$$V_{\theta,b,p}(x) \le \zeta V_{\theta,b,p'}(x),
\quad \forall x \in \X,\ \forall p,p' \in P.$$
Then, for any switching signal with dwell time
$\tau_d > \frac{\ln \zeta}{\kappa}$ the closed-loop switched system $\Xi_g$ is ISS over $\X$.
\end{corollary}
\begin{proof}
By Theorem~\ref{th:guarantee}, for each $p \in P$,
the closed-loop subsystem $\Xi_{g,p}$ admits an ISS-CLF
$V_{\theta,b,p}$ and is ISS over $\X$.
The assumed inter-mode comparison condition ensures
that the family $\{V_{\theta,b,p}\}_{p \in P}$
satisfies the hypothesis of Theorem~\ref{th:dwell-time}.
Therefore, for any switching signal with dwell time
$\tau_d > \ln(\zeta)/\kappa$, Theorem~\ref{th:dwell-time} implies that
the overall switched system $\Xi_g$ is ISS over $\X$.
\end{proof}
\begin{corollary}
Let $V_{\theta,b}$ be a single neural network and
$\{g_{\theta,p}\}_{p\in P}$ be mode-dependent neural controllers.
If the conditions of Theorem~4.6 are satisfied for all $p\in P$
with the \emph{same} Lyapunov network $V_{\theta,b}$,
then the closed-loop switched system is ISS over $\X$
under arbitrary switching.
\end{corollary}

\section{Numerical Simulation}
To validate the proposed neural ISS-CLF-based controller synthesis for switched nonlinear systems, we consider the switched Lotka--Volterra predator--prey model \cite{lotka2002contribution}. For this study, we only consider switching signals under dwell time constraints. 
\begin{equation}
\dot x_1 = ax_1 - bx_1x_2 + g_{\sigma}u,\qquad  
\dot x_2 = -cx_2 + dx_1x_2 + h_{\sigma}u,
\label{eq:predator_prey}
\end{equation}
where the switching signal $\sigma(t)\in\{1,2\}$ selects the mode-dependent input channels 
\[
(g_1,h_1)=(1,0),\qquad (g_2,h_2)=(0,1).
\]
where, $x_1$ denotes the number of prey, while $x_2$ represents the number of predators and $u$ is the human interaction, where $u < 0$ indicates species removal and $u>0$ indicates species introduction. Thus, switching corresponds to alternating human intervention between the prey and predator populations. The objective is to design neural controllers $g_{\theta,p}$ and neural ISS-CLFs $V_{\theta,p}$ that ensure (i) ISS of each subsystem, (ii) forward invariance of the compact set $X$, and (iii) ISS of the overall switched system under an admissible dwell time.
\begin{figure*}
    \centering
    \begin{subfigure}{0.65\textwidth}
        \centering
        \includegraphics[width=\textwidth, height=5 cm]{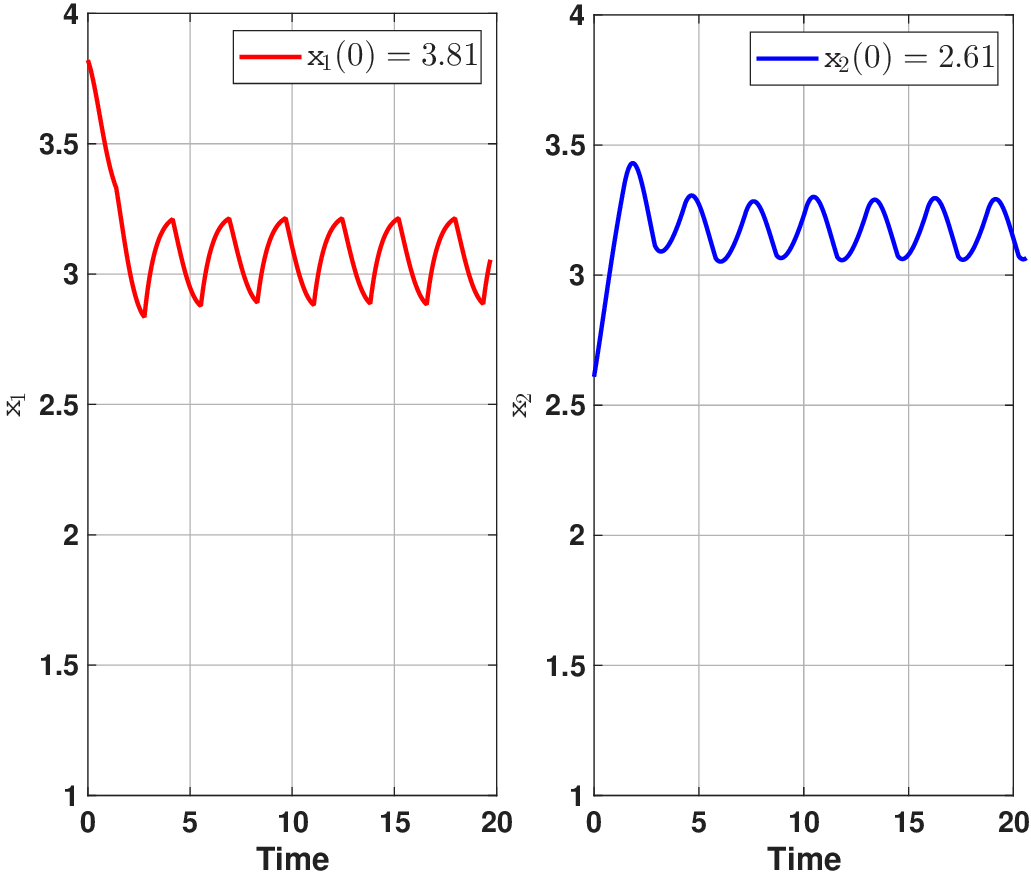}
        \caption{}
    \end{subfigure}
    \hfill
    \begin{subfigure}{0.32\textwidth}
        \centering
        \includegraphics[width=\textwidth, height=5 cm]{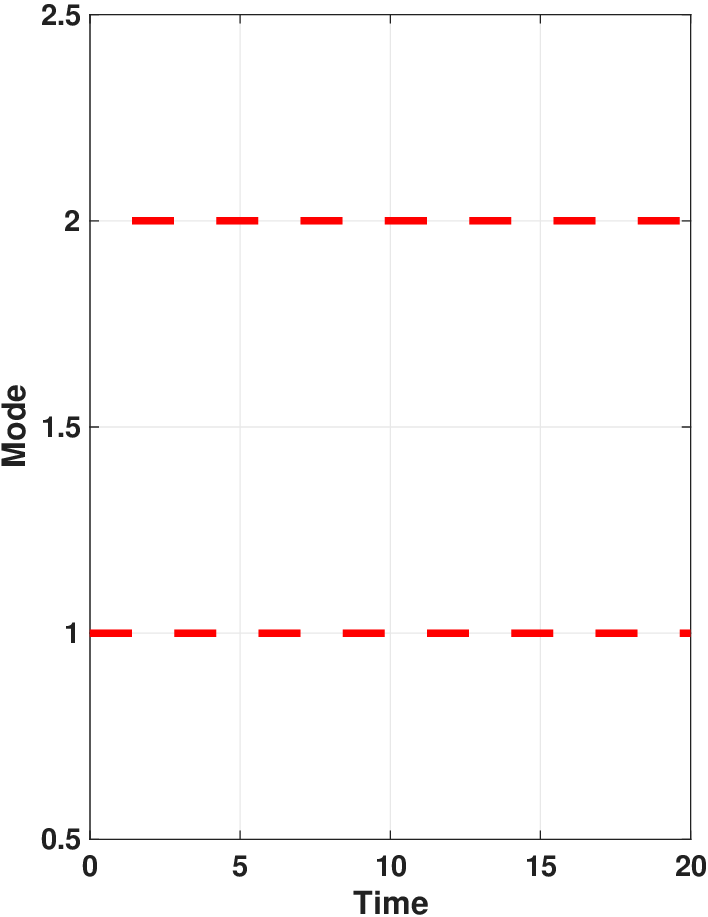}
        \caption{}
    \end{subfigure}
    \vspace{-0.2cm}
    \caption{(a) Evolution of Predator and Prey Population with respect to time corresponding to switched system \eqref{eq:predator_prey}, (b) Switching Sequence.}
    \label{fig:sim1}
\end{figure*}
\subsection{Training Setup}

The compact sets $\X =[1,4]^2\subset\mathbb{R}^2$ and $\W = [0, 1] \subset\mathbb{R}$ are discretized using deterministic $\varepsilon$-covers. For each mode $p\in\{1,2\}$, a neural ISS-CLF $V_{\theta,p}$ and a controller $g_{\theta,p}$ are trained using Algorithm \ref{algo:NN_training}. The losses in \eqref{eq:loss_LF}--\eqref{eq:loss_ineq} enforce the ISS-CLF decrease conditions alongside the invariance of $\X$ through the CBF constraint while satisfying the Lipschitz bounds of the neural networks. Training stops once the verification conditions in Theorem \ref{th:guarantee} are met for both modes.

\subsection{Closed-Loop Simulation}

After training, we simulate the closed-loop dynamics under a switching signal satisfying the dwell-time lower bound derived in Section~\ref{subsec:dwelltime}. Using the initial condition $x(0)=[3.81\;\;2.61]^\top\in \X$, the trajectories shown in Fig.~1 confirm that both states remain inside the safe invariant set $\X$, the neural controller regulates the dynamics toward an ISS under the chosen dwell time.

These results demonstrate the correctness of the proposed neural ISS-CLF framework for switched nonlinear systems with cross-coupled dynamics.

\subsection{Dwell-Time Computation}
\label{subsec:dwelltime}
To ensure input-to-state stability of the overall switched system, Theorem \ref{th:dwell-time} requires the dwell time $\tau_d$ to satisfy $\tau_d > \frac{\ln \zeta}{\kappa}$ where, $\kappa=\min(\kappa_1,\kappa_2)$ is the minimum ISS decay rate over all modes, $\zeta\ge 1$ is the inter-mode Lyapunov comparison constant satisfying $V_p(x)\le \zeta\,V_{p'}(x),\qquad \forall x\in \X,\;\forall p,p'\in\{1,2\}.$ During training, the ISS-CLF derivative constraint $ \frac{\partial V_p}{\partial x} f_p(x,g_p(x,w)) \le -\kappa_p V_p(x) + \sigma_p(|w|) $
is enforced with user-defined $\kappa_p$. Hence,
$\kappa=\min(\kappa_1,\kappa_2)$. After training, a dense grid of points in $X$ is used to evaluate:
$S(x) = \max_{p,p'\in\{1,2\}} \frac{V_p(x)}{V_{p'}(x)}$. The numerical estimate
$\zeta = \sup_{x\in X} S(x)$ and $\kappa$ are then substituted to obtain the minimum dwell time ensuring ISS.  
For example, if $\kappa=0.45$ and $\zeta=1.52$, then
$\tau_d > \frac{\ln(1.52)}{0.45} \approx 0.94\;\text{s}$. In the simulation, we use $\tau_d = 1.5\,$s that guarantees ISS of the switched Lotka--Volterra system.
\section{Conclusion}

This paper developed a neural-network–based control framework for unknown switched nonlinear systems that provides formal guarantees of input-to-state stability (ISS) and safety within compact operating
regions. By learning mode-dependent ISS-CLF and corresponding feedback controllers from data, we established ISS of each closed-loop subsystem together with robust forward invariance of
the prescribed safe set. We then demonstrated that, under an explicit dwell-time requirement, these subsystem-level guarantees lift to ISS of the overall switched system, building on traditional multiple Lyapunov function theory. The proposed framework preserves the rigor of Lyapunov-based switched system analysis while enabling controller synthesis in the absence of explicit system models.
In particular, it recovers classical arbitrary-switching ISS and safety results as a special case when a common neural Lyapunov function exists, thereby unifying common and multiple Lyapunov function approaches within a single learning-based methodology.
The numerical example illustrates how the learned certificates ensure both stability under admissible switching signals. Future research directions include extending the framework to stochastic and state-dependent switching, interconnected and large-scale switched systems, and reducing conservatism in the dwell-time condition through adaptive or average dwell-time analysis.

\bibliographystyle{unsrt} 
\bibliography{reference} 

\end{document}